\documentclass{article}

\usepackage{arxiv}

\usepackage[utf8]{inputenc} 
\usepackage[T1]{fontenc}    
\usepackage{hyperref}       
\usepackage{url}            
\usepackage{booktabs}       
\usepackage{amsfonts}       
\usepackage{nicefrac}       
\usepackage{microtype}      
\usepackage{lipsum}		    
\usepackage{xcolor}
\usepackage{graphicx}
\usepackage{natbib}
\usepackage{doi}
\usepackage{listings}
\usepackage{moreverb}
\usepackage{multicol}        
\usepackage{listings}        

\definecolor{back}{rgb}{0.92,0.92,0.95}
\lstdefinestyle{mystyle}{
    backgroundcolor=\color{back},   
    basicstyle=\ttfamily\footnotesize,
    breakatwhitespace=false,         
    breaklines=true,
    keepspaces=true,                 
    numbers=left,                    
    numbersep=5pt
}

\title{A Hierarchical Approach to exploiting Multiple Datasets from TalkBank}

\author{
    Man Ho Wong\\
	Department of Neuroscience\\
	Department of Linguistics\\
    University of Pittsburgh\\
	Pittsburgh, PA 15260 \\
	\texttt{m.wong@pitt.edu} \\
}

\renewcommand{\headeright}{}

\renewcommand{\undertitle}{Research tools}
\renewcommand{\shorttitle}{A Hierarchical Approach to exploiting Multiple Datasets from TalkBank}

\hypersetup{
pdftitle={A Hierarchical Approach to exploiting Multiple Datasets from TalkBank},
pdfsubject={Applied linguistics},
pdfauthor={Man Ho Wong},
pdfkeywords={Data pipeline, Data mining, TalkBank, CHILDES, Large dataset, Applied linguistics, Corpus linguistics, Natural language processing,  Open science},
}

\begin{document}
\maketitle
\thispagestyle{plain}  

\begin{abstract}

TalkBank is an online database that facilitates the sharing of linguistics research data. However, the existing TalkBank's API has limited data filtering and batch processing capabilities. To overcome these limitations, this paper introduces a pipeline framework that employs a hierarchical search approach, enabling efficient complex data selection. This approach involves a quick preliminary screening of relevant corpora that a researcher may need, and then perform an in-depth search for target data based on specific criteria. The identified files are then indexed, providing easier access for future analysis. Furthermore, the paper demonstrates how data from different studies curated with the framework can be integrated by standardizing and cleaning metadata, allowing researchers to extract insights from a large, integrated dataset. While being designed for TalkBank, the framework can also be adapted to process data from other open-science platforms.
 
\end{abstract}


\begin{multicols}{2}

\section{Introduction}
Recent development in open science platforms has not only greatly increased the transparency of scientific research but also enables global collaboration. These platforms provide access to vast, publicly available datasets, offering the opportunity to gain valuable insights that can only be extracted from a larger sample size. 

TalkBank \citep{MacWhinney2007}, a famous example of such platforms, is a publicly accessible database consisting of datasets contributed by researchers from around the world. The platform covers a diverse range of linguistics topics, including language acquisition, speech–language pathology, and sociolinguistics. While most datasets in TalkBank have been used in published studies, there is likely a substantial amount of hidden information yet to be uncovered from the data. In addition, data from different studies can potentially be integrated into one large dataset to provide a larger sample size, enhancing data reliability and facilitating more robust analyses.

Currently, TalkBank's API \citep{kowalski2019talkbankdb} allows researchers to access the database using custom code. However, the API's data filtering capabilities are limited, hindering comprehensive data exploration. Moreover, the API only allows downloading individual files separately, posing challenges for efficient batch processing of large datasets (Table \ref{compare}).

To overcome the aforementioned limitations of the API, we developed a scalable framework for building pipelines bypassing the API. In the following sections, we will explain the architecture of the framework by building a pipeline to process child speech data from the Child Language Data Exchange System (CHILDES) \citep{doi:10.1177/014272370002006006} in TalkBank. The pipeline built in this paper is available at the following GitHub repository (\texttt{talkbank\_pipeline\_tutorial.ipynb}):\\
\href{https://github.com/manhowong/talkbank-pipeline}{github.com/manhowong/talkbank-pipeline}

 For those interested, a ready-to-use pipeline is also available for adoption to other datasets in TalkBank. (See \texttt{talkbank\_pipeline.ipynb} in the above repository.)

\begin{table*}
	\caption{Comparison between data retrieval using TalkBank's API and the framework presented in this paper.}
    \vspace{5pt}
	\centering
	\begin{tabular}{lll}
		\toprule
		 & Framework in this paper & TalkBank's API \\
		\midrule
		Data filtering & Fully customizable & Available query terms only \\
		File download & Entire corpus/ dataset & Single CHAT file \\
        Data extraction from files & Supported via \texttt{PyLangAcq} package & Supported\\
		\bottomrule
	\end{tabular}
	\label{compare}
\end{table*}

\section{Requirements}

\subsection{Data sourcing}

\paragraph{Source}
The pipeline in this paper retrieves data from the Child Language Data Exchange System (CHILDES) \citep{doi:10.1177/014272370002006006} in TalkBank. Human-annotated recording transcripts in CHAT format were downloaded from CHILDES using the pipeline.

\paragraph{Corpora}
The framework was demonstrated using North American English as the target language due to its extensive, publicly accessible data. There are 47 North American English corpora in CHILDES. Transcripts of child speech were collected from the following 13 corpora in CHILDES: Bates, Bernstein, Brown, Clark, Demetras2, Gleason, HSLLD, Hall, Hicks, Nelson, NewmanRatner, Post and VanHouten. These corpora were selected by the pipeline based on a set of criteria (see \href{https://github.com/manhowong/talkbank-pipeline}{\texttt{talkbank\_pipeline\_tutorial.ipynb}} for details).

\subsection{System}

The initial steps of the framework (scanning download URLs, screening datasets, and downloading data) require a stable internet connection. After the initial steps, data processing is done locally and sufficient storage is required to download the data.

\subsection{Software}
The code was written in Python 3.9.7. For easier demonstration, scripts are organized into Jupyter notebooks. To run the code, a Jupyter Notebook interface is required. You can also run the code on Google Colab.

\paragraph{Python packages}
\texttt{PyLangAcq} (0.16.0) \citep{lee-et-al-pylangacq:2016} is required to read the CHAT files. In addition, the following packages are also required: \texttt{Pandas}, \texttt{bs4}, \texttt{NumPy}, \texttt{Requests}, \texttt{Urllib}, \texttt{Zipfile}, and \texttt{tqdm} (optional, for showing progress bar during running).

\section{Framework overview}

\begin{figure*}
	\centering
	\includegraphics[width=\textwidth]{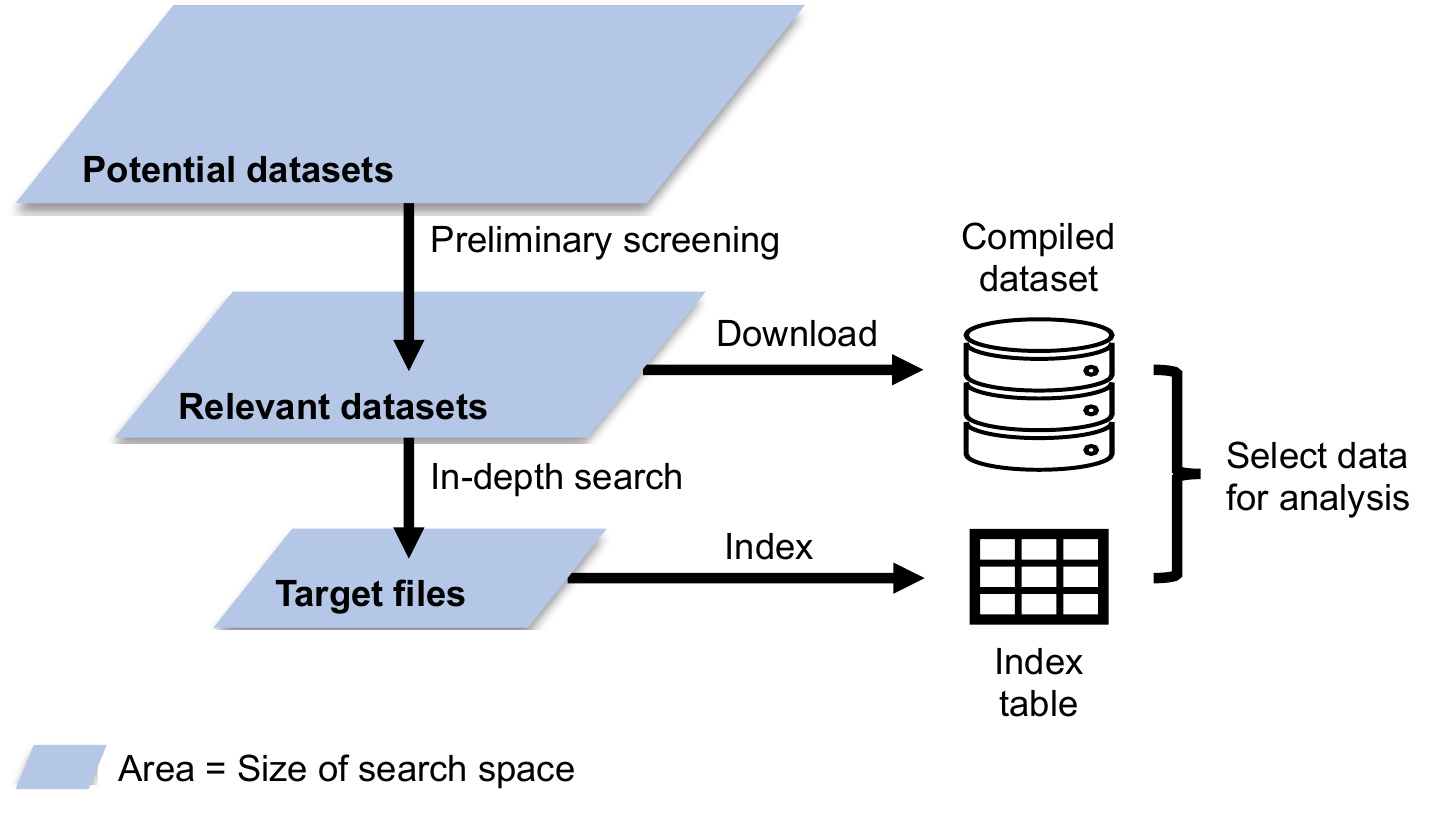}
    \caption{Framework architecture. A hierarchical approach is implemented to narrow down the search space progressively with increasingly stringent criteria at each level. Relevant corpora are downloaded after preliminary screening. Target files are indexed at the last level. The compiled dataset can be accessed offline via the index table.}
    \label{framework}
\end{figure*}

The framework starts with defining the potential datasets and identifying the source URLs (Figure \ref{framework}). Users need to specify the datasets they want to include in the search for the data they will need. For more efficient data processing, the search space is narrowed down progressively using a hierarchical searching strategy where each level of the hierarchy employs more stringent criteria. 

At the first level of the hierarchy, the framework performs a preliminary screening of relevant datasets that may contain the desired data. This initial screening does not employ stringent criteria nor inspect every file (see next section for implementation), as the goal here is just to reduce the search space. The identified corpora are then downloaded as a single dataset to the local drive. This compiled dataset facilitates data access during subsequent stages of the framework and data analysis, eliminating the need to repeatedly access the remote data source.

At the second level of the hierarchy, the framework performs an in-depth search with more specific criteria to further refine the search space. In addition, every file is inspected throughout the search. Files that match the specified criteria are indexed and stored in a table. This index table serves as a reference for locating the target data within the large dataset compiled in the previous step. Using the index table, researchers can quickly access the target data required for analysis without repeating data searching in the remote data source.

\section{Demonstration: Building a pipeline to retrieve child speech data from TalkBank}

In this demonstration, our goal is to retrieve the data containing child speech or child-directed speech recorded from typically developed children of age 0 to 72 months. In addition, we are only interested in files that contain information about the participant's socio-economic status. Such complex data selection criteria can not be implemented with TalkBank's API.

To implement the above criteria, we will build a Python pipeline using the framework presented in this paper. We will use the pipeline to search for the files we need and compile a large dataset by integrating data from multiple North American English (Eng-NA) corpora in the CHILDES collection of TalkBank. The compiled dataset will be indexed with metadata from file headers.

The pipeline includes the following major steps:
\begin{enumerate}
	\item Define potential dataset(s) and identify source URLs 
    \item Screen for relevant corpora
    \item Download relevant corpora to the local drive
    \item Search for target files and index them
    \item Standardize header labels
    \item Add participant identifier (optional)
\end{enumerate}

\subsection{Define potential dataset(s) and identify source URLs}

TalkBank maintains a well-organized directory structure, where each collection, such as CHILDES or ASDBank, has its dedicated directory. Within each collection, the datasets are stored under the "data" subdirectory:

\begin{verbatim}
TalkBank
|-----Collection 1
|-----Collection 2
|-----Collection 3 (e.g. CHILDES)
      |-----Data
            |-----dataset 1
            |-----dataset 2
            |-----dataset 3 (e.g. Eng-NA)
                  |-----Corpus1.zip
                  |-----Corpus2.zip
                  |-----Corpus3.zip
\end{verbatim}

For example, the North American English dataset (Eng-NA) in the CHILDES collection can be accessed through the following URL, where the subdomain corresponds to the collection's name (i.e. CHILDES):

\begin{center}
	\url{https://childes.talkbank.org/data/Eng-NA}
\end{center}

Corpus data are stored as zip files within each dataset. To get the download URLs for the dataset(s), you will first need to specify the TalkBank collection you are interested in, such as CHILDES. You may also provide the name(s) of the specific dataset(s) you want to download. You can visit the TalkBank Browser to look up the directory name of a collection or a dataset: Select the collection in the pull-down menu and navigate to the dataset you are interested in. The name of the collection and/or dataset is indicated in the directory path under the pull-down menu. Alternatively, go to the collection's data page, e.g. \url{https://childes.talkbank.org/data/} for CHILDES.

The pipeline scans the collection's data subdirectory and automatically fetches the URLs for all the downloadable zip files you will need for the next step. Note that this pipeline retrieves download URLs for public collections only. For password-protected collections, please follow the instructions in TalkBank's documentation.

\subsection{Screen for relevant corpora}
\label{screen}

For cases where not all the corpora found in the potential datasets are needed, an optional step can be taken to screen for relevant corpora based on the CHAT files required. This step not only reduces the total download size but also narrows the search space for target CHAT files down the line. This step is especially important if you have a large set of potential datasets.

\begin{figure*}  
\centering
\begin{lstlisting}[caption={Example of a nested if-conditional statement (written in pseudocode) using metadata in file headers to inspect CHAT files in CHILDES.}, label={ifcondition}]
if `CHI' exists in `participants' field of header
    and(
        `ses' field of `CHI' in `participants' field of header is not empty
        
        or (`MOT' exists in `participants' field of header and 
        `ses' field of `MOT' in `participants' field of header is not empty)
        
        or (`MOT' exists in `participants' field of header and 
        `education' field of `MOT' in `participants' field of header is not empty)
    )
\end{lstlisting}
\end{figure*}

The pipeline employs the \texttt{PyLangAcq} package \citep{lee-et-al-pylangacq:2016} to scan the metadata in the header of each CHAT file to check whether the file satisfies the screening criteria. For efficient screening, no stringent criteria will be used in this step, but rather simple criteria that are sufficient enough to narrow down the search space. In addition, the pipeline does not inspect every CHAT file in a corpus thoroughly but instead moves on to the next corpus as soon as it finds a file satisfying the screening criteria. This approach significantly improves screening efficiency. The download URLs for the corpora containing at least one file matching the screening criteria will be returned.

Screening criteria can contain any metadata available in the file header. Listing \ref{ifcondition} shows an example of an if-conditional statement screening for CHAT files that involve a child participant ('CHI') and contain info about either the child's socio-economic status (SES), the mother's ('MOT') SES, or the mother's education. In this example, the pipeline will return a list of download URLs for the corpora containing at least one file that satisfies the if-condition.

\subsection{Download relevant corpora to the local drive}

The zip file of every corpus found in the previous step will be downloaded and extracted to the local drive. Each zip file contains all the CHAT files in a corpus. Instead of downloading the entire corpus, individual CHAT files can be accessed through TalkBank's official API so that one can download only the CHAT files they need. However, in most cases, sending a request and downloading one CHAT file at a time takes a much longer time than downloading the entire corpus. 

\subsection{Search for target files and index them}

After downloading the data, the pipeline will inspect the metadata of every CHAT file to search for files that meet the user-defined criteria. Similar to Step \ref{screen}, the criteria can contain any metadata available in the file header.

The output of the pipeline -- a table indexing the target CHAT files based on their metadata -- is generated in this step (e.g. Table \ref{index}). This index table will be used to select CHAT files in future data analysis. Each row of the index table corresponds to a CHAT file (i.e. an entry), and each column represents a file header field. Available file header fields are documented in the CHAT manual. Note that not all files contain the same set of header fields. To determine what header fields are available in a file, you can either view the header of the file using the TalkBank Browser or download the file and view it on your computer.

\begin{table*}[]
	\caption{Index table generated by the pipeline. Each entry corresponds to a CHAT file.}
    \vspace{5pt}
	\centering
    \begin{tabular}{llllllllll}
    \toprule
      & file\_path   & corpus & participants & name          & age\_m & sex    & group & ses & study\_type \\
    \midrule
    0 & ../amy.cha   & Bates  & CHI, MOT & Target\_Child & 20.0   & female & TD    & MC  & cross, toyplay, TD \\
    1 & ../betty.cha & Bates  & CHI, MOT & Betty         & 20.0   & female & TD    & MC  & cross, toyplay, TD \\
    2 & ../chuck.cha & Bates  & CHI, MOT & Chuck         & 20.0   & male   & TD    & MC  & cross, toyplay, TD \\
    3 & ../doug.cha  & Bates  & CHI, MOT & Doug          & 20.0   & male   & TD    & MC  & cross, toyplay, TD \\
    ... & & & & & & & & &\\
    
    \bottomrule
    \end{tabular}
    \label{index}
\end{table*}

\subsection{Standardize header labels}

Since the target CHAT files may come from different studies, CHAT files recorded under the same experimental conditions might be indexed differently with different header values (i.e. "header labels"). For example, either TD or typical could be used to label files from typically developed children (Table \ref{labelExamples}). In addition, since metadata might be entered by researchers manually, human errors such as typos or missing data could be found. To ensure data integrity and consistency, the index table undergoes a cleaning process before being saved as a Python pickle file for future use. The cleaning process involves standardizing header values, removing any incorrect values, handling missing values, and resolving any inconsistencies or discrepancies within the table. 

To clean the index table, you may want to start by inspecting the header labels first. To get all the labels used for a header field (corresponds to a column in the table), you can call the function \texttt{get\_labels(<COLUMN NAME>)}. If you are not sure what each of the labels represents, you can find the information about the experimental conditions as well as the study design on the homepage of the corpus where the CHAT file belongs to. The Jupyter Notebook \texttt{talkbank\_pipeline\_tutorial.ipynb} demonstrates how header labels can be standardized and how missing values can be handled according to the documentation of the corpus involved.

\begin{table*}
	\caption{Example of a header field before standardization. Different labels are used in the header field `group' for the same condition across different corpora, such as `typical' and `TD', or `MOT\_Older' and `MOT\_other'.}
    \vspace{5pt}
	\centering
	\begin{tabular}{ll}
		\toprule
		Corpus & Labels for the header field `group'  \\
		\midrule
        Bates & `TD' \\
        Bernstein & `unspecified', `TD' \\
        Brown & `unspecified', `TD' \\
        Clark & `TD' \\
        Demetras2 & `unspecified', `TD' \\
        Gleason & `normal', `unspecified', `typical', `TD' \\
        HSLLD & `unspecified' \\
        Hall & `unspecified', `White,UC', `TD' \\
        Hicks & `unspecified' \\
        Nelson & `unspecified' \\
        NewmanRatner & `TD' \\
        Post & `TD' \\
        VanHouten & `MOT\_Adolescent', `MOT\_Adolescent\_', `MOT\_Older', `MOT\_Older\_',\\
        & `MOT\_adolescent', `MOT\_older', `TD', `unspecified' \\
		\bottomrule
	\end{tabular}
	\label{labelExamples}
\end{table*}

\subsection{Add participant identifier (optional)}

In each corpus, participants can be identified by their names or IDs within the corpus. However, in situations where data from different corpora are pooled together, there is no unique identifier available for participants across multiple corpora. For example, a participant named "Adam" can be found in two corpora (Brown and VanHouten) in CHILDES, though they are clearly different people. In addition, some participants may be associated with multiple CHAT files, such as those involved in longitudinal studies where repeated observations are made with the same participants over a period of time. A unique identifier becomes essential when conducting individual-based analyses or when tracking participant-related information across different datasets or studies.

One straightforward approach to creating a participant identifier is to combine the participant's name with the name of the corpus to which the CHAT file belongs. By concatenating these two pieces of information, a unique identifier can be generated for each participant. This combined identifier ensures that participants with the same name in different corpora or with multiple CHAT files can be distinguished and analyzed individually.

\section{Summary}

This paper presents a novel framework that addresses the limitations of TalkBank's API in data selection and batch processing. The framework enables customizable data filtering, allowing researchers to use complex criteria for retrieving relevant data from multiple corpora/ datasets. Using a hierarchical search approach, the framework reduces the search space sequentially and makes data processing more efficient. Moreover, the framework generates an index table that facilitates future data analysis by eliminating the need to search for the same datasets repeatedly.

While the framework is demonstrated using the CHILDES data in TalkBank as an example, it is applicable to other remote databases as well. Written in Python, the framework can readily be adapted to the RAPIDS ecosystem for high-performance GPU processing of big data. The framework presented in this paper opens up possibilities for researchers to efficiently exploit multiple datasets from open science platforms like TalkBank and similar resources.

\section{Acknowledgement}
I would like to thank Prof. Na-Rae Han (Robert Henderson Language Media Center/ Department of Linguistics, University of Pittsburgh) for providing valuable input during the development of this project.

\bibliographystyle{unsrtnat}
\bibliography{references}
\end{multicols}

\end{document}